\documentclass[aps,twocolumn,amsmath,amssymb,showpacs,prl]{revtex4}
\usepackage{epsf}

\newcounter{saveeqn}%
\newcommand{\alpheqn}{\setcounter{saveeqn}{\value{equation}}%
\stepcounter{saveeqn}\setcounter{equation}{0}%
\renewcommand{\theequation}{\mbox{\arabic{saveeqn}\alph{equation}}}}%
\newcommand{\reseteqn}{\setcounter{equation}{\value{saveeqn}}%
\renewcommand{\theequation}{\arabic{equation}}}%
\newcommand{\ts}{\textstyle }
\newcommand{\lsim}{ {{}_{\ts <} \atop {\ts \sim}} }
\newcommand{\gsim}{ {{}_{\ts >} \atop {\ts \sim}} }

\begin{document}

\title{Dispersion Anomalies in Bilayer Cuprates and the Odd Symmetry
of the Magnetic Resonance}
\author{M. Eschrig$^{1}$ and M. R. Norman$^{2}$}
\affiliation{$^1$Institut f{\"u}r Theoretische Festk{\"o}rperphysik,
Universit{\"a}t Karlsruhe, 76128 Karlsruhe, Germany\\
$^2$Materials Science Division, Argonne National Laboratory, Argonne,
Illinois 60439}
\date{June 27, 2002}
\begin{abstract}
We present a theoretical model which accounts for recent 
angle resolved photoemission data in bilayer
cuprate superconductors.
Lineshapes and dispersions of the various bonding and antibonding
features in the spectra are quantitatively reproduced.
The observed dispersion anomalies are consistent
with the interaction of electrons with a bosonic mode which is
odd with respect to the layer indices, a unique property of the magnetic
resonance observed by inelastic neutron scattering.
\end{abstract}
\pacs{74.25.Jb, 74.72.Hs, 79.60.Bm}

\maketitle

Recent angle resolved photoemission (ARPES) experiments on
bilayer cuprate superconductors were able for the first time
to resolve a bilayer splitting between bonding and antibonding 
bands \cite{Feng01,Gromko02,Kordyuk02,Kaminski02}.
The dispersion near the $(\pi,0) $ point of the Brillouin zone
shows an unusual asymmetry between bonding and antibonding 
self energy effects.  In particular,
Feng {\it et al.} \cite{Feng01} found that the energy distribution curves 
(EDCs, taken
at constant momentum) in overdoped Bi$_2$Sr$_2$CaCu$_2$O$_{8+\delta}$ 
($T_c=65$K) consist of three features: an antibonding band (AB)
peak near 20 meV, 
a bonding band (BB) peak near 40 meV, and a
bonding hump near 105 meV.  Recently, Gromko {\it et al.} \cite{Gromko02}
reported strong self energy effects in the 
dispersions derived from momentum distribution curves 
(MDCs, taken at constant energy) in similar samples ($T_c=58$K).
Near momentum $(k_x,k_y)=(1,0.13)\pi/a$, an $S$-shaped
dispersion anomaly, discussed previously in Ref.~\cite{Norman01},
was shown to be present only in the bonding band MDC,
at binding energies between 40 meV and 60 meV.  In both experiments,
a low energy double peak structure in the EDC was only resolvable in the
same momentum region.

In this letter, we demonstrate that all of these features can be
explained by a model
which assumes that low energy scattering of electrons 
between the bonding and antibonding bands is strong
compared to scattering within each of those bands.
As scattering events which connect different bilayer bands
are odd with respect to permutation of the layers within a bilayer, this
implies that the corresponding bosonic excitations which mediate
such scattering must be dominant in the odd channel.
The magnetic resonance observed in inelastic neutron scattering has
exactly this property \cite{Rossat91,Fong99}, 
and moreover has the correct energy to reproduce the observed dispersion
anomalies. 
Thus, the recent ARPES experiments on bilayer cuprates provide
independent support for a strong coupling between this 
resonance and electronic excitations.

If the electrons are phase coherent between the two planes, then
the spectra will exhibit separate
bonding ($b$) and antibonding ($a$) features with (normal state)
dispersions given by $\xi_{\vec{k}}^{(b)} $ and $\xi_{\vec{k}}^{(a)} $. 
In the cuprates, the resulting
energy splitting is anisotropic \cite{Chak93,Feng01}
\begin{equation}
\label{eq1}
\xi_{\vec{k}}^{(a)}-\xi_{\vec{k}}^{(b)}=
\frac{1}{4}t_{\perp } (\cos k_x  - \cos k_y )^2.
\end{equation}
In the superconducting state, the dispersions are modified by the presence
of the $d$-wave order parameter given by
$\Delta_{\vec{k}}=\Delta_0 (\cos k_x - \cos k_y )$. 
In agreement with experiment \cite{Feng01,Kordyuk02},
we assume $\Delta_0$ to be the same for the bonding and antibonding bands.
Then, the dispersion in the 
superconducting state takes the form
$(E^{(a,b)}_{\vec{k}})^2=
{(\xi^{(a,b)}_{\vec{k}})^2+(\Delta_{\vec{k}})^2}$.

This ``non-interacting'' picture, though, is insufficient to
describe the three observed dispersion features.  We are able to account
for them in a model where electrons are coupled to a resonant spin
mode situated below a gapped spin fluctuation continuum.
Such a spectrum is observed in
inelastic neutron scattering experiments \cite{Rossat91}.
In bilayer materials, the spin susceptibility is a matrix in the 
layer indices, having elements diagonal 
($\chi_{aa}$, $\chi_{bb}$) and off-diagonal $(\chi_{ba}$, 
$\chi_{ab}$) in the bonding-antibonding representation. 
The components of the spin susceptibility transforming as
even and odd with respect to the layer indices are given by
$\chi_e=\chi_{aa}+\chi_{bb}$ and $\chi_o=\chi_{ab}+\chi_{ba}$.
For identical planes, $\chi_{aa}=\chi_{bb}$ and $\chi_{ab}=\chi_{ba}$.
The measured susceptibility is then given by
\begin{equation}
\chi=\chi_e \cos^2 \frac{q_zd}{2} + \chi_o \sin^2 \frac{q_zd}{2}
\end{equation}
where $d$ is the separation of the layers within a bilayer.
The resonance part, $\chi_{res}$, was found to exist only
in the odd channel, whereas the continuum part, $\chi_{c}$, enters in
both \cite{Rossat91}. 
Thus, $\chi_{o}=\chi_{res}+\chi_{c}$ and $\chi_{e}=\chi_{c}$.
This means the resonance mode can only scatter electrons between
the bonding and antibonding bands. In contrast,
the spin fluctuation continuum scatters both within and between
these bands.  As we demonstrate below, the odd symmetry of the resonance
is crucial in reproducing the ARPES spectra.

We employ for the single layer self energy the functional form of 
Ref.~\cite{Eschrig00}.
It is proportional to the convolution of the 
Gor'kov-Green's function $\hat G$ with
the dynamic spin susceptibility $\chi $.
We write this self energy symbolically as
$\hat \Sigma =g^2\chi \ast \hat G$
(the hat denotes the 2x2 particle hole space, and $g$ is the coupling
constant).
For a bilayer system, there is a separate self energy for each band,
$\hat \Sigma^{(a,b)}$.
Using the above notation, these self energies are given by
\alpheqn
\begin{eqnarray}
\label{bilayer1}
\hat\Sigma^{(b)} &=& \frac{g^2}{2}\left\{ \chi_{res} \ast \hat G^{(a)} + 
\chi_{c} \ast \left( \hat G^{(b)}+\hat G^{(a)} \right) \right\}
\\
\label{bilayer2}
\hat\Sigma^{(a)} &=& \frac{g^2}{2}\left\{ \chi_{res} \ast \hat G^{(b)} +
\chi_{c}\ast \left( \hat G^{(b)}+\hat G^{(a)} \right) \right\}.
\end{eqnarray}
\reseteqn
Dispersion anomalies arise mainly from
coupling to the resonance mode. This means that dispersion anomalies
in the bonding band
are determined by the antibonding spectral function and vice versa.
(The spectral functions for the bonding and antibonding bands are given by 
the imaginary parts of the diagonal components of the Green's functions.)
Because the antibonding band is (in contrast to the bonding band)
close to the chemical potential at
$(\pi,0)$ \cite{Feng01}, the associated van Hove singularity
leads to a larger self energy for the bonding band.

\begin{figure}
\centerline{
\epsfxsize=1.7in{\epsfbox{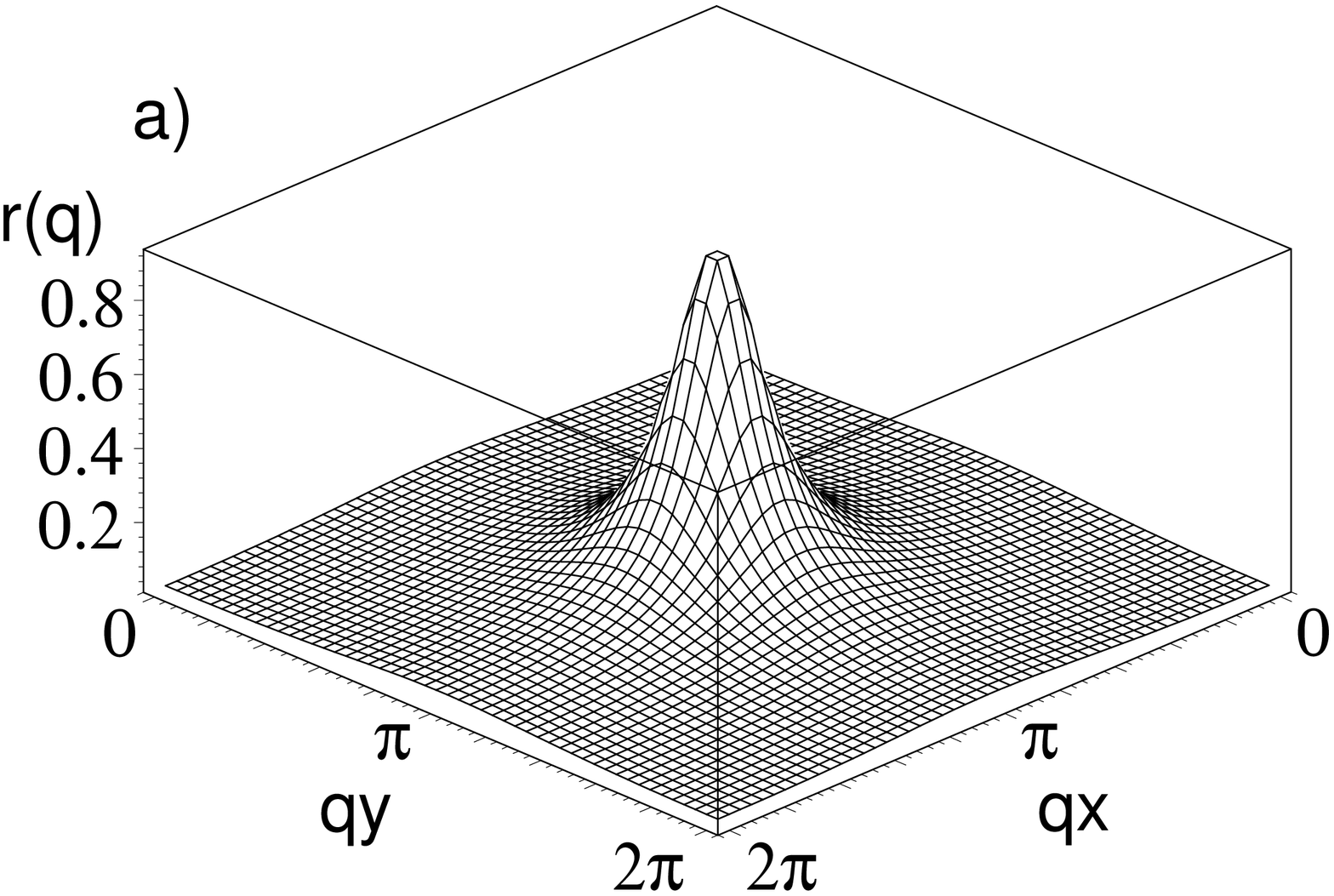}}
\epsfxsize=1.7in{\epsfbox{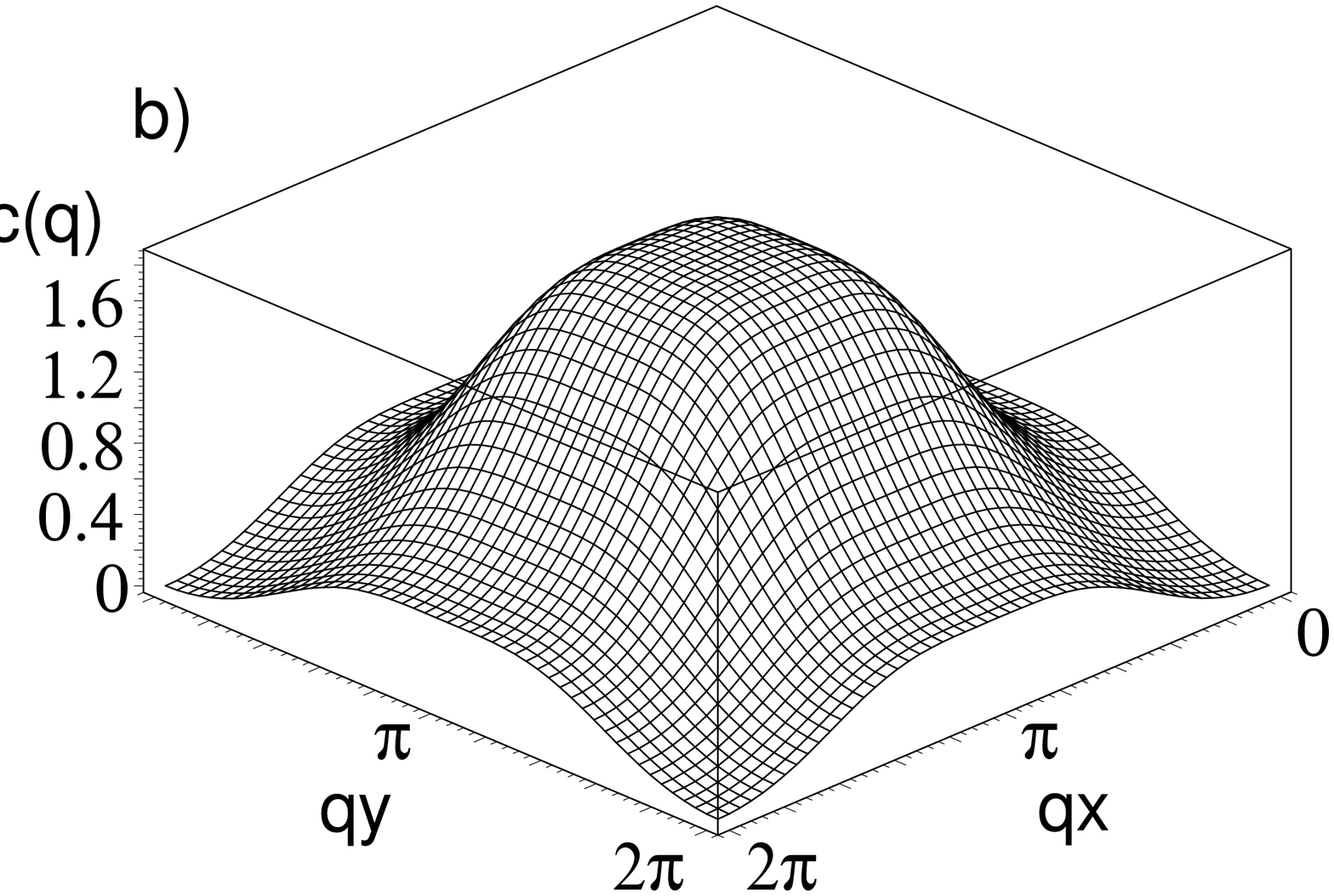}}
}
\caption{ \label{fig0}
Momentum dependence of a) the resonance mode
and b) the gapped spin fluctuation continuum.
The resonance mode is peaked at $\vec{Q}=(\pi,\pi)$ with a 
correlation length equal to twice the lattice constant, $\xi_{res} = 2a$.
The continuum spectrum, in contrast, is rather broad around $\vec{Q}$.
}
\end{figure}

The self energy effects can be
described very accurately by using a model spin spectrum which consists of
a perfectly sharp resonance mode below a gapped continuum with gap
$\Delta_h$. We approximate the intensity of the continuum to be constant.
In reality, the intensity of the continuum decays at 
high energies. However, as we are interested in self energy effects 
in an energy range $|\epsilon | < 200$ meV, 
the above approximation is adequate.
A high frequency cut-off in the convolution integrals in 
Eqs. (\ref{bilayer1}) and (\ref{bilayer2}) was introduced.
The precise cut-off procedure however does not affect the low energy physics.
Any variation in the cut-off can be accounted for by a readjustment of 
the coupling constants and
the normal state renormalization factor. 
We model the resonance mode, 
$R=2{\rm Im}\chi_{res}$,
and the continuum, 
$C=2{\rm Im}\chi_c$, by
\alpheqn
\begin{eqnarray}
R_{\omega, \vec{q}} &=& 2 r(\vec{q}) \left\{
\delta (\omega-\Omega_{res}) - \delta (-\omega - \Omega_{res}) \right\},
\\
C_{\omega, \vec{q}} &=& 2 c(\vec{q}) \left\{
\Theta (\omega-2\Delta_{h}) - \Theta (-\omega - 2\Delta_{h}) \right\}.
\end{eqnarray}
\reseteqn
where the $\Theta $ function is zero for negative argument and one otherwise.
The momentum dependences of the resonance mode and the continuum,
given by the functions $r(\vec{q})$ and $c(\vec{q})$, are plotted in 
Fig.~\ref{fig0}.
The resonance mode, shown in 
Fig.~\ref{fig0}a, is peaked at $\vec{Q}$, with a correlation length of
$\xi_{res}=2a $, where $a$ is the lattice constant. The gapped continuum,
shown in Fig.~\ref{fig0}b, is much broader.
This is motivated by the experimental data \cite{Rossat91}, which show that the
continuum is enhanced around $\vec{Q}$ with a correlation length of only
0.5 lattice constants. Also, the momentum dependence
of the continuum excitations exhibit experimentally 
a flat behavior around $\vec{Q}$, as in Fig.~\ref{fig0}b.
To simplify the model, we use the same functional form for the
gapped continuum in the even and odd scattering channels. This is consistent
with the superconducting state data,
where the gap in the odd channel (about twice the maximal superconducting gap), 
is close to the optical gap in the even channel.

We have chosen parameters appropriate for the overdoped sample
($T_c=65 $K) studied in experiment \cite{Feng01}. 
The normal state dispersion is obtained from
a 6 parameter tight-binding fit to experimental data \cite{Norman95},
with the bilayer splitting given by Eq.~\ref{eq1}.
In order to determine the total of 7 parameters for this fit, we used
three Fermi wavevectors,
$\vec{k}_N=(0.37,0.37) \pi /a $,  $\vec{k}_{A}^{(b)}=(1,0.217)\pi /a$, and
$\vec{k}_{A}^{(a)}=(1,0.135)\pi /a$, 
the Fermi velocity at $\vec{k}_N$, $\vec{v}_N=2.316 $ eV\AA , and
the normal state band energies at $(\pi,0)$, 
$\xi_M^{(b)}=-105 $ meV and $\xi_M^{(a)}=-18 $ meV
(the corresponding bilayer splitting is 
$\xi_M^{(a)}-\xi_M^{(b)}=t_{\perp}=87 $ meV).
Finally, to obtain the correct overall shape of the dispersion, we 
fixed the band energy at $(\pi,\pi)$, using
a reasonable value $\xi_{Y}=0.8 $ eV. 
The high energy ($|\epsilon| \gsim 200$ meV)
dispersions are not affected strongly
when going from the normal to the superconducting state.
However, even in the normal state, the bare dispersion is
renormalized by the normal state spin fluctuation continuum.
To account for this renormalization, we multiply the above dispersion by
a factor of 1.4 in order to obtain the bare dispersion.
For the remaining parameters of the model, we use 
$\Delta_0=16$ meV \cite{Feng01},
$\Omega_{res}=27$ meV, $\Delta_h=0.9\Delta_0$, $g^2r(\vec{Q})=0.15 $eV$^2$,
and $g^2c(\vec{Q})=0.72 $eV.
The value for the resonance energy was obtained from the 
relation $\Omega_{res}=4.9 k_BT_c$ found experimentally to hold for
overdoped 
Bi$_2$Sr$_2$CaCu$_2$O$_{8+\delta}$ \cite{Rossat91,Fong99,Zasadzinski01}.
The resonance weight $r(\vec{Q})$ has not been measured for overdoped
materials. For optimally doped Bi$_2$Sr$_2$CaCu$_2$O$_{8+\delta}$, it is
$0.95 \mu_B^2$ per plane \cite{Fong99}, and we estimated for this case
$g=0.65 $ eV \cite{Eschrig00}.
Assuming for the overdoped sample the same coupling
constant, the above value
for $g^2r(\vec{Q})$ would imply a resonance weight of $0.36 \mu_B^2$ per plane.
Similarly, with this coupling constant and
our value for $g^2c(\vec{Q})$, we obtain a (2D)-momentum averaged 
continuum contribution of $1.7 \mu_B^2/$eV per plane
(gotten by summing the even and odd channels 
for energies $\omega \lsim 0.2 eV$). 
Our calculations are for a temperature $T=10 $K.
Note we use unrenormalized Green's functions in Eqs.~\ref{bilayer1} and
\ref{bilayer2}. This approximation is sufficient to explain a large variety 
of data, and can be justified by considerations discussed in Ref.~\cite{Vilk97}.

\begin{figure}
\centerline{
\begin{minipage}{1.4in}
\epsfxsize=1.4in{\epsfbox{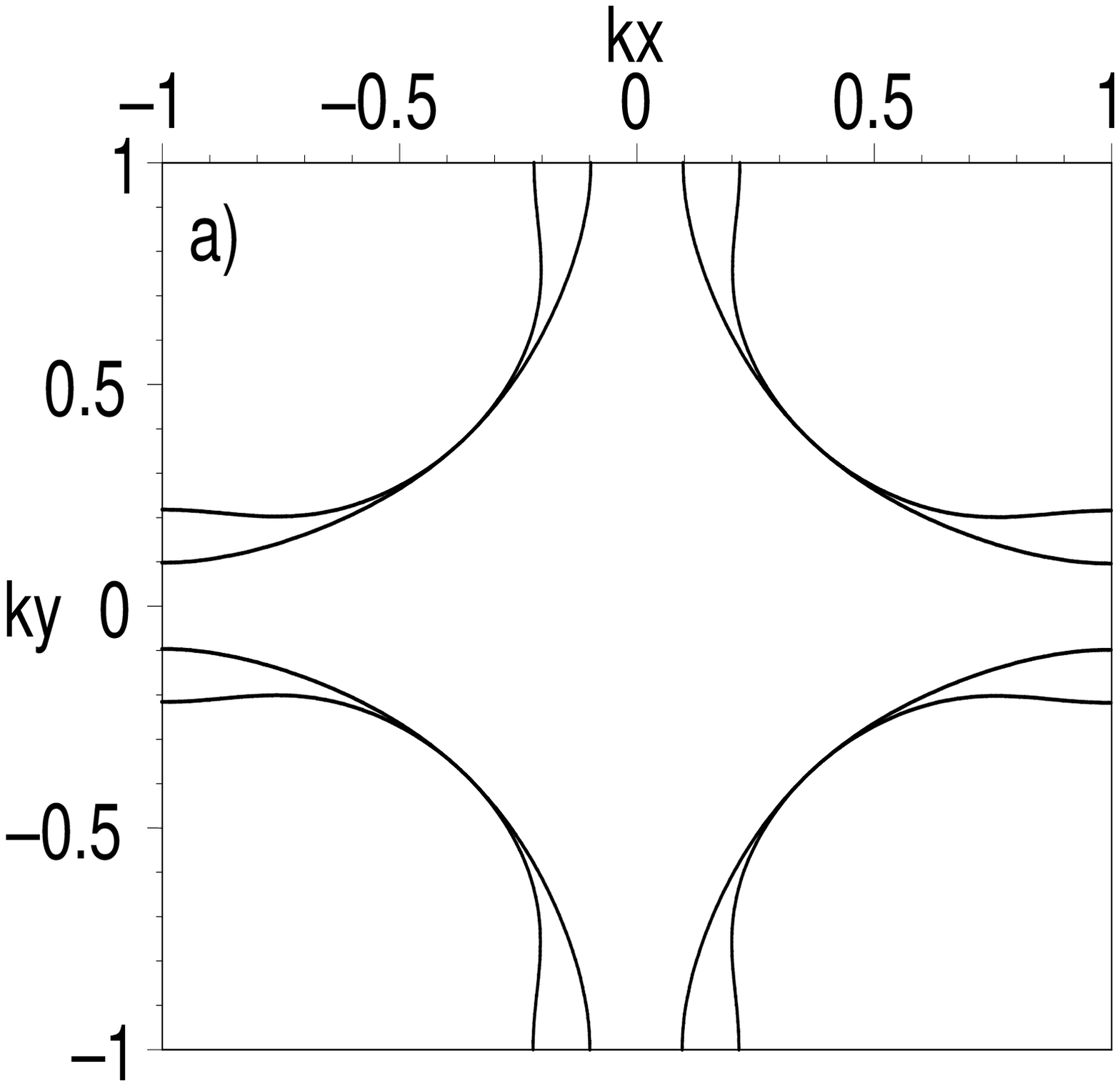}}
\epsfxsize=1.4in{\epsfbox{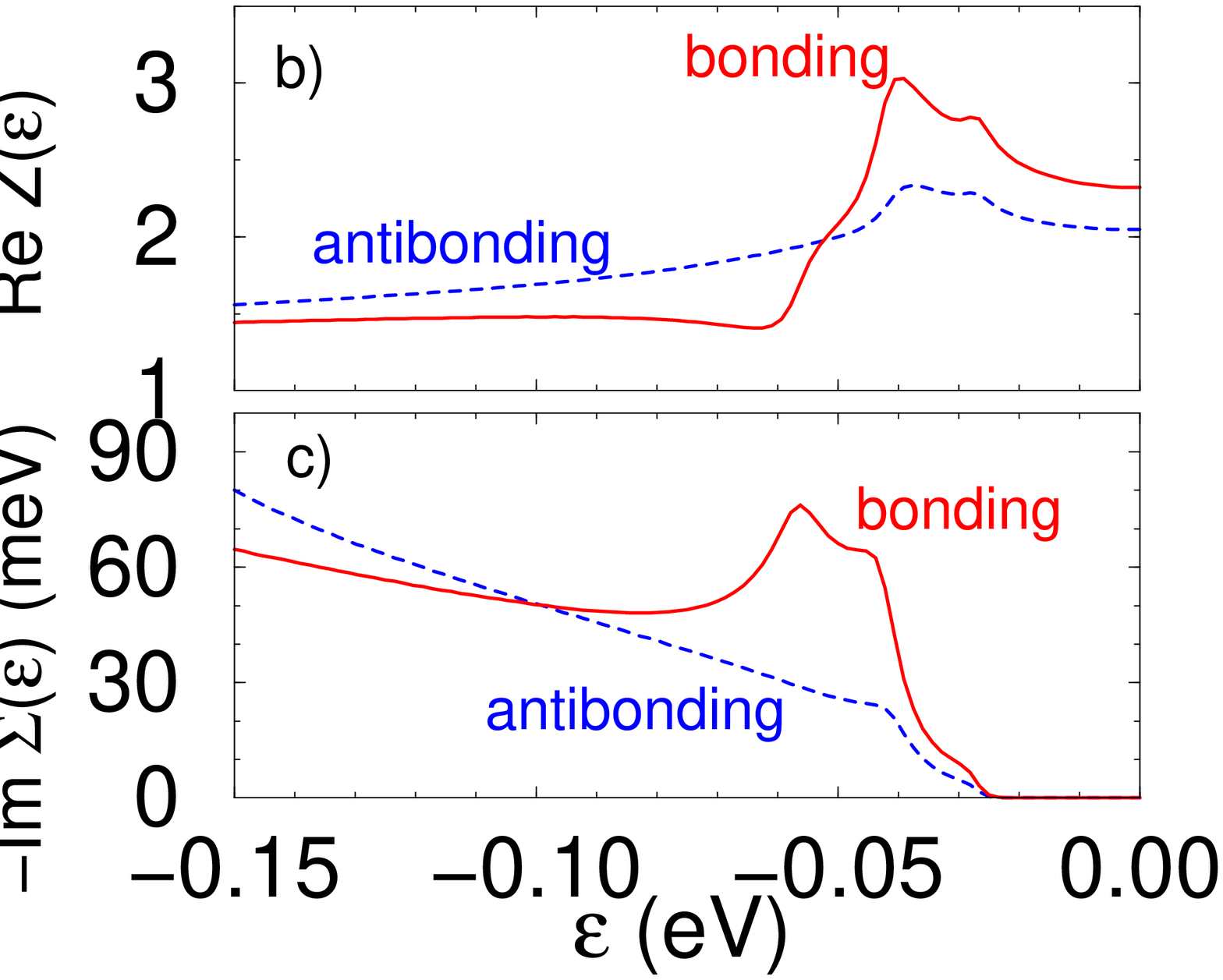}}
\vspace{0.3in}
\end{minipage}
\begin{minipage}{1.7in}
\epsfxsize=1.7in{\epsfbox{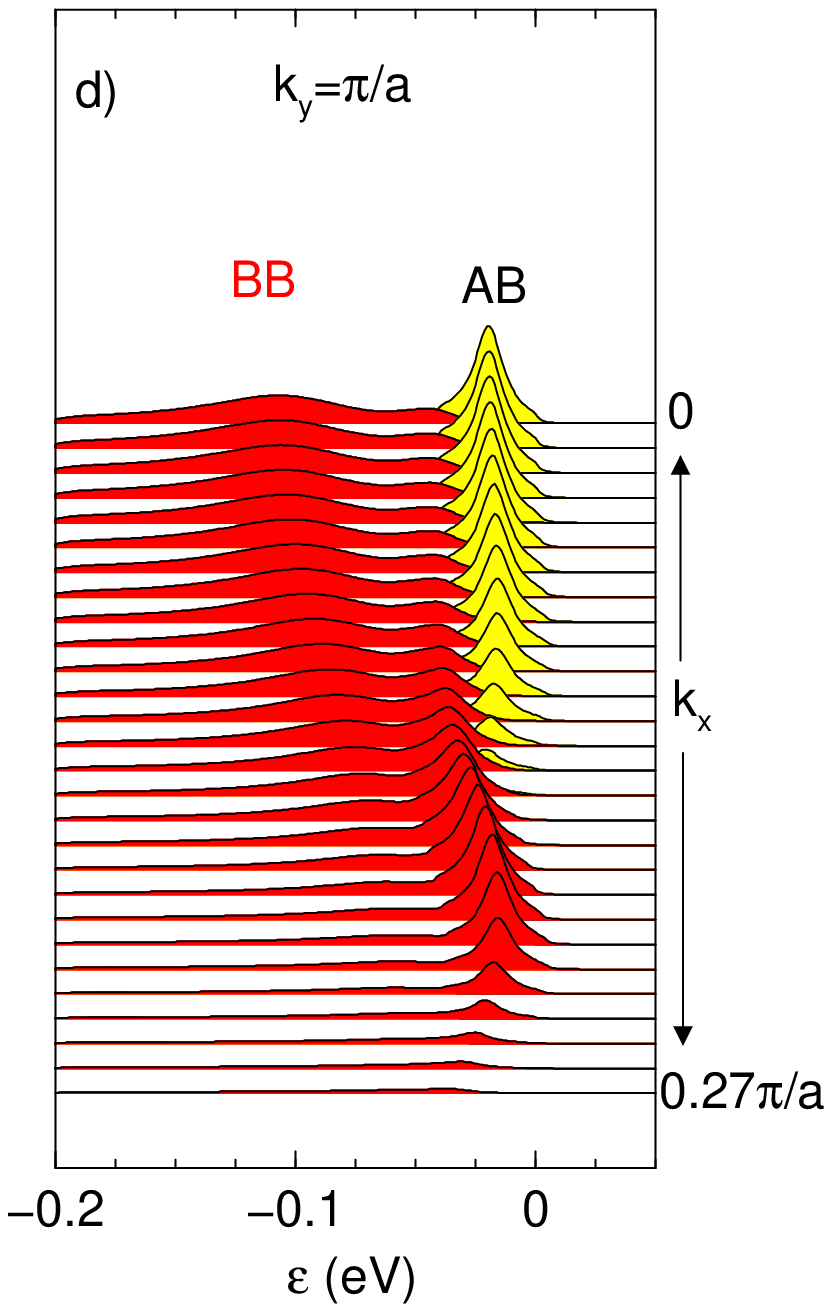}}
\end{minipage}
}
\caption{\label{fig2}
a) Tight binding Fermi surfaces for antibonding and bonding bands
in overdoped Bi$_2$Sr$_2$CaCu$_2$O$_{8+\delta}$ ($T_c=65 $K).
b) Renormalization function and c)
imaginary part of the self energy at
the $(\pi,0)$ point of the zone
for the bonding band (BB) and antibonding band (AB).
d) Spectral functions for $k_y=\pi /a$, and $k_x $ varying
from $0$ to $0.27 \pi /a$. 
For comparison with experiment,
the spectra are convolved with a
Lorentzian energy resolution function (FWHM 12 meV).
}
\end{figure}

The bonding and antibonding normal
state Fermi surfaces are shown in Fig.~\ref{fig2}a. The bilayer splitting
is maximal near the $(\pi,0)$ points
of the zone. Thus, we will concentrate on this region
in the following.
In Fig.~\ref{fig2}b we show the renormalization function for bonding and
antibonding bands at the $(\pi,0)$ point.
The renormalization is stronger for the bonding band
than the antibonding band. This is a result of the
proximity of the antibonding saddle point singularity to the chemical potential.
As is seen in this figure as well, both bands are renormalized up to high 
energies. 
The high energy renormalization approaches that of the normal state
dispersion (1.4).

The imaginary part of the self energy is shown in Fig.~\ref{fig2}c
for the bonding and antibonding bands. 
As emission processes
are forbidden for $|\epsilon| < \Omega_{res}$, the
imaginary part of the self energy is zero in this range.
Due to scattering events to the antibonding band, electrons
in the bonding band have a large imaginary part of the self energy
in the range between 40 and 60 meV. These events
are dominated by emission of the resonance, and
are enhanced due to the van Hove singularity
in the antibonding band close to the chemical potential.
In contrast,  the imaginary part of the antibonding self energy 
is not enhanced because the bonding band is far away from the chemical potential
at $(\pi,0)$. Consequently, it shows linear behavior over a
wide energy range, with a gap at low energies 
($|\epsilon| < \Omega_{res}$).

Fig.~\ref{fig2}d presents the intensities for the bonding and antibonding
spectra. 
The antibonding spectra consist of a low energy AB peak, and the bonding
spectra have a low energy BB peak and a higher energy BB hump feature.
In agreement with experiment (\cite{Feng01} and
\cite{Gromko02}), the width of the EDC 
spectrum is large for the BB hump, but not so for the BB and AB peaks.
We also mention that the BB peak is well defined in the whole region
between the BB Fermi crossings on either side of $(\pi,0) $,
but the finite energy resolution 
does not allow to resolve it near $(\pi,0)$ as seen in Fig.~\ref{fig2}d.

\begin{figure}
\centerline{
\epsfxsize=3.4in
\epsfbox{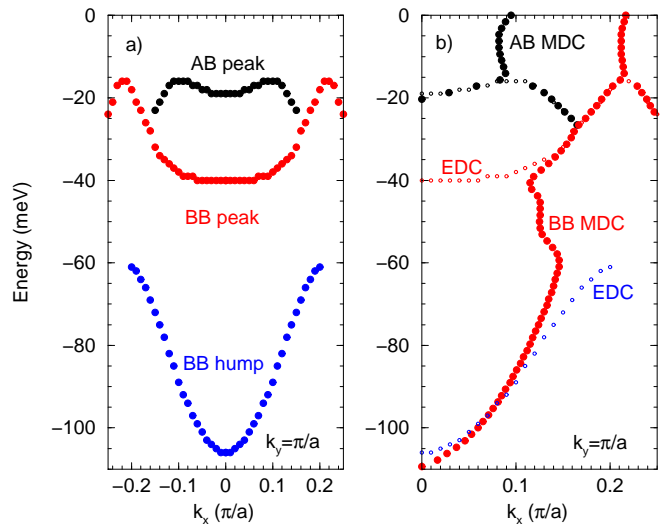}
}
\caption{ \label{fig1}
Dispersion of EDC peak positions (left) and MDC peak positions (right)
near the $(\pi,0)$ point of the Brillouin zone. The EDC dispersion consists
of three branches, one antibonding peak, one bonding peak and one bonding
hump. The MDC bonding dispersion shows a characteristic $S$ shape 
behavior.
The small symbols in the right picture show again the EDC peak positions 
for convenience.
}
\end{figure}
In Fig.~\ref{fig1} we show our results for dispersion of 
the EDC and MDC peak positions. In the EDC dispersions, Fig.~\ref{fig1}a,
we reproduce the experimentally observed three branches \cite{Feng01}, one
antibonding peak and two bonding branches, denoted 
`BB peak' and `BB hump'. The BB peak has a very
flat dispersion near $k_x=0$ in agreement with experiment \cite{Feng01}.
Its position at 40 meV is approximately given by $\Omega_{res} + \Delta_A$,
where $\Delta_A$ is the gap at the antibonding Fermi crossing.
Thus, the energy separation between the AB peak at the AB Fermi crossing 
and the BB peak at $(\pi,0)$
is a measure of the resonance mode energy $\Omega_{res}$ in
overdoped compounds. The BB hump position at high binding
energies (105meV) is determined by the normal state dispersion
of the bonding band.
Because the spin fluctuation continuum changes only at low energies when
going from the normal to the superconducting state, the position
of the BB hump maximum is not very different from the normal state
BB dispersion. This is in agreement with experiment \cite{Feng01}.
The intensity of the AB peak decreases quickly when it approaches
the BB peak, but is strong at $(\pi,0)$ because of
the proximity of the AB band to the chemical potential in this region.

In Fig.~\ref{fig1}b, we present the MDC dispersions
(for comparison we also reproduce the EDC dispersions as small symbols).
The MDC dispersion consists of two branches, an AB MDC branch and
a BB MDC branch. The self energy effects are most clearly observable
in the BB MDC branch. In the binding energy range between 40meV and 60meV,
there is an $S$-shaped `break' region, connecting the BB hump EDC branch
with the BB peak EDC branch. This $S$-shaped behavior reproduces the finding
of recent experiments \cite{Gromko02}.

The dispersion anomalies observed in the {\it bonding} band are a mirror of
the large number of states close to the chemical potential 
near $(\pi,0)$ for the {\it antibonding} band.
Scattering events involving a mode with energy $\Omega_{res}$ couple
the bonding band electrons in the energy region between 40 and 60 meV
strongly to those antibonding band electrons.
The corresponding processes are in the odd scattering channel.
The energy range of anomalous dispersion is shifted by the 
resonance mode energy with respect to the antibonding binding energies.

\begin{figure}
\centerline{
\epsfxsize=2.8in{\epsfbox{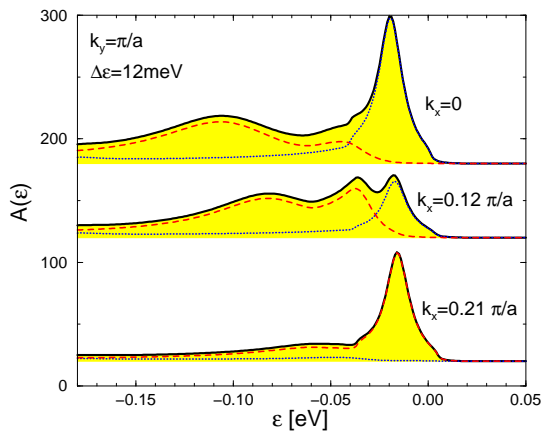}}
}
\caption{\label{fig3}
Spectral functions for $k_y=\pi /a$ at $k_x=0$, $k_x=0.12 \pi /a $ and
$k_x=0.21 \pi /a$. The full lines are the sum of antibonding (dotted)
and bonding (dashed) spectral functions. 
For comparison with experiment, a
Lorentzian energy resolution function of 12 meV width was assumed.
The double peak structure
is clearly resolved for $k_x=0.12 \pi /a$, as in experiment 
\cite{Feng01,Gromko02}.
}
\end{figure}

Finally, in
Fig.~\ref{fig3} we compare spectra for three positions in the Brillouin
zone, corresponding to the spectra presented in Refs.~\cite{Feng01,Gromko02}.
For each spectrum, the bonding (dashed) and antibonding (dotted) contributions
are indicated.
The spectra are convolved with a Lorentzian energy resolution function to
allow for direct comparison with experiment.
We reproduce all experimental findings. First, at $(\pi,0)$,
only the BB hump and the AB peak are resolved. This is due to resolution
effects mentioned above.
Second, near the AB Fermi crossing, the spectra show a characteristic
double peak structure, with a relatively sharp
AB peak and a BB peak separated from a broad BB hump. Third, at the
BB Fermi crossing, only the BB peak is observed. The BB hump is so small
in intensity that it only leads to a kink-like feature in the spectrum.

We have presented a theory to account for the experimentally
observed self energy effects in the bilayer split bands in
double layer high temperature superconductors. We reproduced quantitatively
the EDC dispersions, the MDC dispersions, and the spectral lineshapes.
We found that the ARPES data are consistent with the interaction of the
electrons with a sharp bosonic mode which is odd in the layer indices,
a property unique to the magnetic resonance observed by inelastic
neutron scattering.

This work was supported by the U. S. Dept. of Energy, Office of Science,
under Contract No. W-31-109-ENG-38.

\end{document}